# Electron transfer channel in the sugar recognition system assembled on nano gold particle

Takayuki Goto[1*], Takeshi Hashimoto[1], Kai Sato[1], Yukihiro Kitamoto[1], Takashi Hayashita[1], Satoshi Iguchi[2], Takahiko Sasaki[2], Dita Puspita Sari [3,4] and Isao Watanabe[4]

*[1]Faculty of Science and Technology, Sophia University, Tokyo 102-8554, Japan,*
*[2]Institute for Material Research, Tohoku University, Sendai 980-8577, Japan,*
*[3]SIT Research Laboratories, Shibaura Institute of Technology, Saitama 337-8570, Japan,*
*[4]Meson Science Laboratory, RIKEN, 2-1 Hirosawa, Wako, Saitama 351-0198, Japan.*

Existence of 1D spin diffusion in the electrochemical sugar recognition system consisting of a nano-sized gold particle (GNP), a ruthenium complex and a phenylboronic acid was investigated by NMR and μSR. When sugar molecules are recognized by the phenylboronic site, the response of electrochemical voltammetry of the Ru site changes, enabling the system to work as a sensitive sugar-sensor. In this recognition process, the change in the electronic state at the boron site caused by sugar must be transferred to the Ru site via alkyl chains. We have utilized the muon-labelled electrons method and the proton NMR to find out a channel of the electron transfer from the phenylboronic acid site to the gold nano particle via the one dimensional alkyl chain. If this transfer is driven by diffusive spin channel, characteristic field dependence is expected in the longitudinal spin relaxation rate of μSR and $^{1}$H-NMR. We have observed significant decrease in the spin relaxation rates with increasing applied field. The result is discussed in terms of low dimensional spin diffusion.

## 1. Introduction

The recently reported electrochemical sugar recognition system consisting of a gold nano-sized particle (GNP), a ruthenium complex and a phenylboronic acids (Fig. 1) attracts much interest because of its high sensitivity for various sugars such as D-glucose or D-fructose[1)]. As sugars are essential biological molecules for human body, simple and prompt measurement of sugar concentration in aqueous solution is a very important subject. So far, various kinds of sugar sensors had been developed. Among them, the fluorescent boronic acids have been focused as chemical sensors due to their high stability and selectivity; such as the internal charge transfer (ICT) system and the photoinduced electron transfer (PET) system.[2-6)] In 2014, another new type of boronic-acid-based sugar sensor, shown in the title, was developed by our colleagues Hashimoto and Endo[1) ] . When sugar molecules are "recognized" by the phenylboronic site of this sensor, boronic ester is formed and simultaneously the hybridization of boron changes from *sp*2 to *sp*3. An excess charge brought by this change is believed to be transferred to the ruthenium complex site and give the drastic response of electrochemical voltammetry, enabling the system to work as a highly sensitive sugar-sensor.[1)]. In this recognition process, it seems to be likely that the change in the electronic state at the phenylboronic acids site is transferred to the ruthenium complex site via 1D alkyl chains. However, microscopic view of its route is still not clear until now.

If this idea of 1D propagation along alkyl chains is correct, it can be detected by measuring the muon-spin relaxation process, which is caused by the magnetic interaction between the muon spin and the moving electron produced by muon itself. That is, the injected positive muon can pick-up one electron to form a neutral atomic state, which is soon thermalized and bonded to a relatively-reactive site on the chain. In other words, in place of recognition process of sugar molecules by phenylboronic acid site, one utilizes muon as a trigger to induce a moving electron and also as an observer of it, which is known as the muon-labelling measurement.[7-10)] This method was first applied to the polyacetylene and later to various systems including polaron motion in oganic polymer chains, or a typical one dimensional system, DNA[7-10)].

The purpose of this study is to find out a possible electron transfer channel from the phenylboronic acid site to the ruthenium complex via gold nano particle. By its findings, we will be given a better understanding of the sugar-recognition mechanism, and also, a possibility to develop a sensor with still higher sensitivity or with more functions[11)].

In μSR, the characteristic dimensionality of the electron motion can be readily found by measuring the LF dependence of the dynamical relaxation rate λ; for one-dimensional motion, it is expected to be proportional to $H_{\rm LF}^{-1/2}$, where $H_{\rm LF}$ is the externally applied field along the muon spin polarization[7-10)]. Here, its field dependence reflects the functional form of Fourier spectrum of the local field fluctuation, generated by diffusive motion of spins. The same field dependence is expected to be observed by NMR, that is, the longitudinal nuclear spin relaxation rate $1/T_1$ [12-16)]. For these two measurements have different target frequency ranges, we expect to obtain comprehensive results by applying both the two techniques in the present study.

## 2. Experimental

The gold nano particles (GNPs) were prepared from $HAuCl_4$ by the reduction of sodium citrate following the general method[1,4)]. The average diameter of the GNPs was determined by the dynamic light scattering (DLS) to be approximately 12.0(2) nm [4]. For the complete sugar sensor, GNPs are attached with both ruthenium complex and phenylboronic acid molecules. However, in the present work, we omitted the former on purpose, because the trivalent Ru ion may dominates the nuclear and muon spin relaxation by its paramagnetic moment via the dipole-dipole interaction. So, we prepared samples with only phenylboronic acid attached and concentrated on the spin diffusion along the alkyl chain connecting GNP and phenylboronic acid.

We prepared two samples with different alkyl chain length zero (B0) and 12 (B12). [1,4,17] The number of phenylboronic acid molecues attached to a GNP is roughly estimated to be 100(20), from the change in the number of Ru complex for Ru-only sample and that assembled with both Ru and phenylboronic acid [17]. The details of sample synthesis and of drying method are described in the previous paper [17].

The longitudinal (LF) field-μSR measurements under the field between 0.04 and 0.38 T were performed at room temperature on a powder sample at Riken-RAL Muon facility using a spin-polarized pulsed surface-muon ($\mu^+$) beam with a momentum of 27 MeV/c. We utilized the asymmetry for the measure of the muon spin polarization [18,19]. The muon spin depolarization data were analyzed with the function $A_0 e^{-\lambda t} + A_1$, where $A_0$ and $A_1$ are the relaxing and baseline asymmetry, respectively, and λ is the depolarization rate due to the dynamical spin fluctuation. Note that in this field region, the effect of quasi-static nuclear spins, that is, Kubo-Toyabe function can be neglected. We first fitted roughly all the relaxation data by the function, and obtained the averaged value of $A_0$, and then with keeping $A_0$ as the fixed averaged value, the fitting was performed on all the data again to obtain the other parameters.

$^1$H-NMR measurements were performed by the conventional spin-echo method with the 6 T cryogen-free magnet. Spectra were obtained by recording the echo amplitude against the external field. The longitudinal spin relaxation rate $1/T_1$ was obtained by recording the echo amplitude against the measurement-repetition cycle. The repetition cycle was extended until the difference between the nuclear spin magnetization and its thermal equilibrium value becomes less than one percent. For the B12 sample, the observed recovery curve of nuclear spin magnetization was fitted with a single exponential function, while the B0 sample, the stretched exponential function $e^{-(t/T_1)^\beta}$ was utilized [20,21].

## 3. Results and Discussion

In Fig. 2, muon-spin-depolarization curves under various fields are shown. Observed

Lorentz type depolarization indicates the existence of dynamical spin fluctuation. The field dependence of λ is shown in Fig. 3, where one can see that λ decreases with increasing $H_{\mathrm{LF}}$ until the highest field 0.38 T. As for the magnitude of depolarization rate, an appreciably high value of λ within the totally non-magnetic compound indicates the existence of some strong relaxation process [7-10], which we believe to be the spin diffusion process as described later.

Next, one may note that initial asymmetry at $t = 0$ increases slowly with increasing applied field until its maximum 0.38 T. This increase is an artifact brought by the applied LF fields, so called the alpha value effect, and is not like what was observed in trans-polyacetylene[7]. The observed its monotonic dependence on $H_{\mathrm{LF}}$ as well as the increasing rate 10 %/T well agree with the facility data of ISIS-RAL, assuring that our interpretation is correct. This also assures that the muon spin relaxation has only a single component in entire field region.

In Fig. 4 and 5, we show NMR data for the B12 sample. The former shows the typical recovery curves of longitudinal nuclear spin relaxation under various fields. One can clearly see that the relaxation becomes slower under higher fields. The two spectra shown in the inset taken under different applied fields indicate that the line width is determined by the homogeneous broadening, and hence assures the sample homogeneity. Obtained relaxation rate $1/T_1$ from recovery curves was plotted against the applied field in Fig. 5, where one can see that $1/T_1$ obeys inverse square root law in the wide field range between 2 and 6 T. Though the magnitude of relaxation rate seems to be very small, that is, the order of a second, its value is comparable to the result of NMR on polyacetylene, where the 1D spin diffusion is concluded [12]. For the sample B0, where phenylboron-acid molecules are directly attached to GNP, $1/T_1$ has no field dependency, as shown in Fig. 6.

Here, we compare the μSR and NMR data to discuss the possibility of 1D spin diffusion. Since in 1D diffusion case, the characteristic field-dependency of $1/T_1 \propto 1 / \sqrt{H}$ can be observed only when the Larmor frequency of muon or nuclear spin is in between the intra- and inter-chain diffusion rates[9,22,23]. Note that here the diffusion rate is defined as $D/a^2$, where $D$ and $a$ are the conventional diffusion constant and the electron hopping step distance, respectively.

The measurement frequency regions of μSR and NMR in the present work are $\gamma_\mu H =$ 10-50 MHz and $\gamma_{\mathrm{H}} H =$ 90-260 MHz, where $\gamma_\mu$, $\gamma_{\mathrm{H}}$ and $H$ are the gyromagnetic ratio of muon, that of proton, and the applied field. The observed results can be consistently explained, if we assume that the inter-chain diffusion rate, that is, 2D or 3D diffusion, is lower than 50 MHz, comparable to μSR frequency, and that the intra-chain diffusion rate along the alkyl chain is higher than 260 MHz, the upper point of NMR frequency[12,13]. With this assumption, one can conclude that the 1D spin diffusion exists, and that the saturating behavior at low fields of μSR data is due to the slow electron motion hopping from one alkyl chain to the other.

Here we add several comments on the interpretation of observed data to validate the above arguments. First, we ignored an extremely fast relaxation with an extremely small fraction seen in the short time region before 1 μs, which can be seen in the 0.05 T data in Fig. 2. This is because no systematic dependence of its fraction on $H_{\mathrm{LF}}$ can be seen. That is, in higher fields, it only looks like a wiggling rather than a fast relaxation. Furthermore, the clear observation of $1/T_1 \propto 1 / \sqrt{H}$ in NMR suggests the existence of spin diffusion, which is, in general, slow, and is expected to modulate only long tail part of the relaxation curves. One can also deny the idea that the reduction of λ at higher-fields above 0.2 T in Fig. 3 corresponds to the fast relaxation model and hence should be treated with the Redfield equation as in the case of paramagnetic systems[24,25]. If that is the case, much faster and field independent NMR relaxation should be observed, which disagrees with our observation.

Finally, we refer to the extremely slow NMR relaxation rate. As shown in Fig. 5, $T_1$ is order of one second, while that of μSR is of micro second. Even if we consider the difference in their gyromagnetic ratios $\gamma_\mu$ and $\gamma_{\mathrm{H}}$, still, the former is more than three orders of magnitude longer. This is straightforward consequence of the fact that in NMR measurements, there is nothing to trigger the spin diffusion, but only a small number of paramagnetic impurities, while in μSR, muons themselves do[14-16].

In order to confirm that the observed behavior of λ comes from the 1D-diffusion along alkyl-chain, investigation on wider region of field, that is, higher than 0.5 T for μSR, and lower than 2 T for $^1$H-NMR is inevitable, which is now on the progress. The comparison of μSR data on B12 with that of B0 will be also of help for the confirmation of the propagation of spin diffusion along alkyl chains. For the full understanding of mechanism of sugar recognition, the role of GNP that connects alkyl chains is very important, and for this purpose, Au-NMR study on the GNP in this system is also planned.

**Summary**

We have investigated the one dimensional spin diffusion on the alkyl-chain in the gold nano particle-based sugar sensing composite by NMR and μSR. For the sample with the longer alkyl chain, the significant field dependence of $1/\sqrt{H}$ was observed in NMR-$T_1$, suggesting the existence of the spin diffusion motion along the chain.

**Acknowledgment**

This work was supported by JSPS KAKENHI Grant Number 21K03452, 23K04792 and 21H01955. A part of this work was performed under the Inter-University Cooperative Research Program of the Institute for Materials Research, Tohoku University.

*E-mail: gotoo-t@sophia.ac.jp

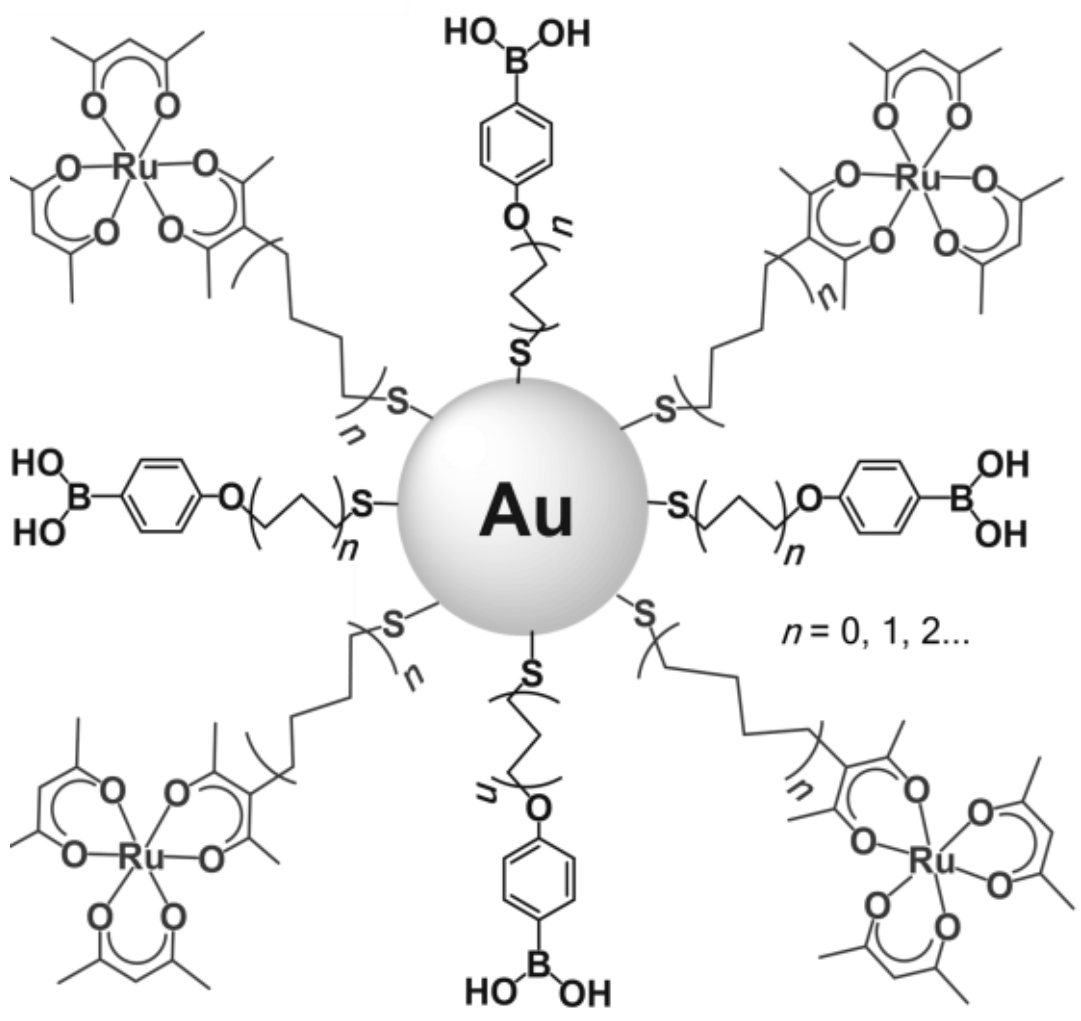

Fig. 1. Schematic drawing of the sugar recognition composite system, consisting of Au nano particle, phenylboronic acid and Ru complex. The index n shows the length of alkyl chain. In the present work, samples with only phenylboronic acid were investigated (see text.).

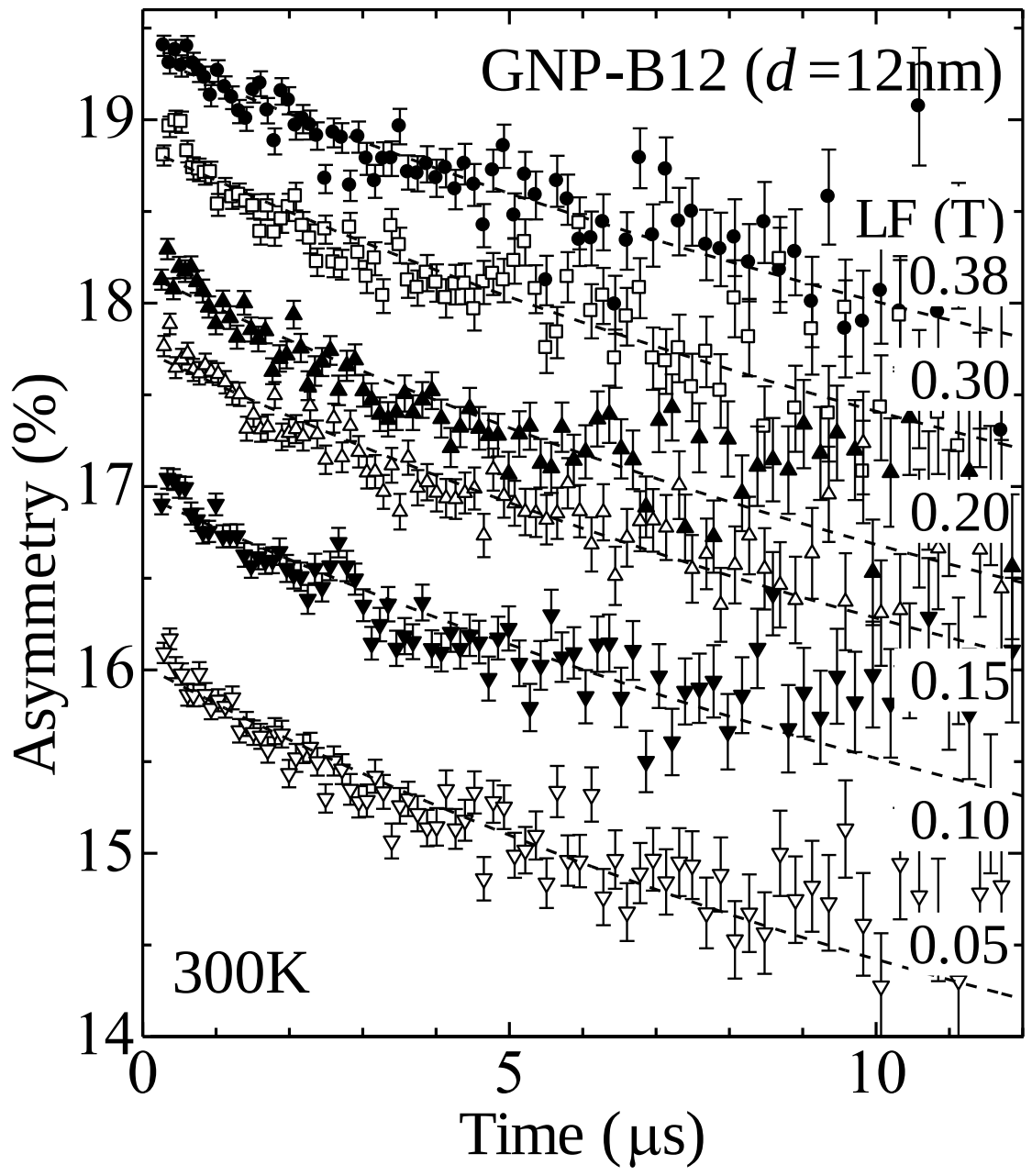

Fig. 2. Typical depolarization curves of μ-SR taken under various longitudinal fields (LF) for the sample B12. No vertical offset in asymmetry is set to each data. Dashed curves show the exponential function fitted to data.

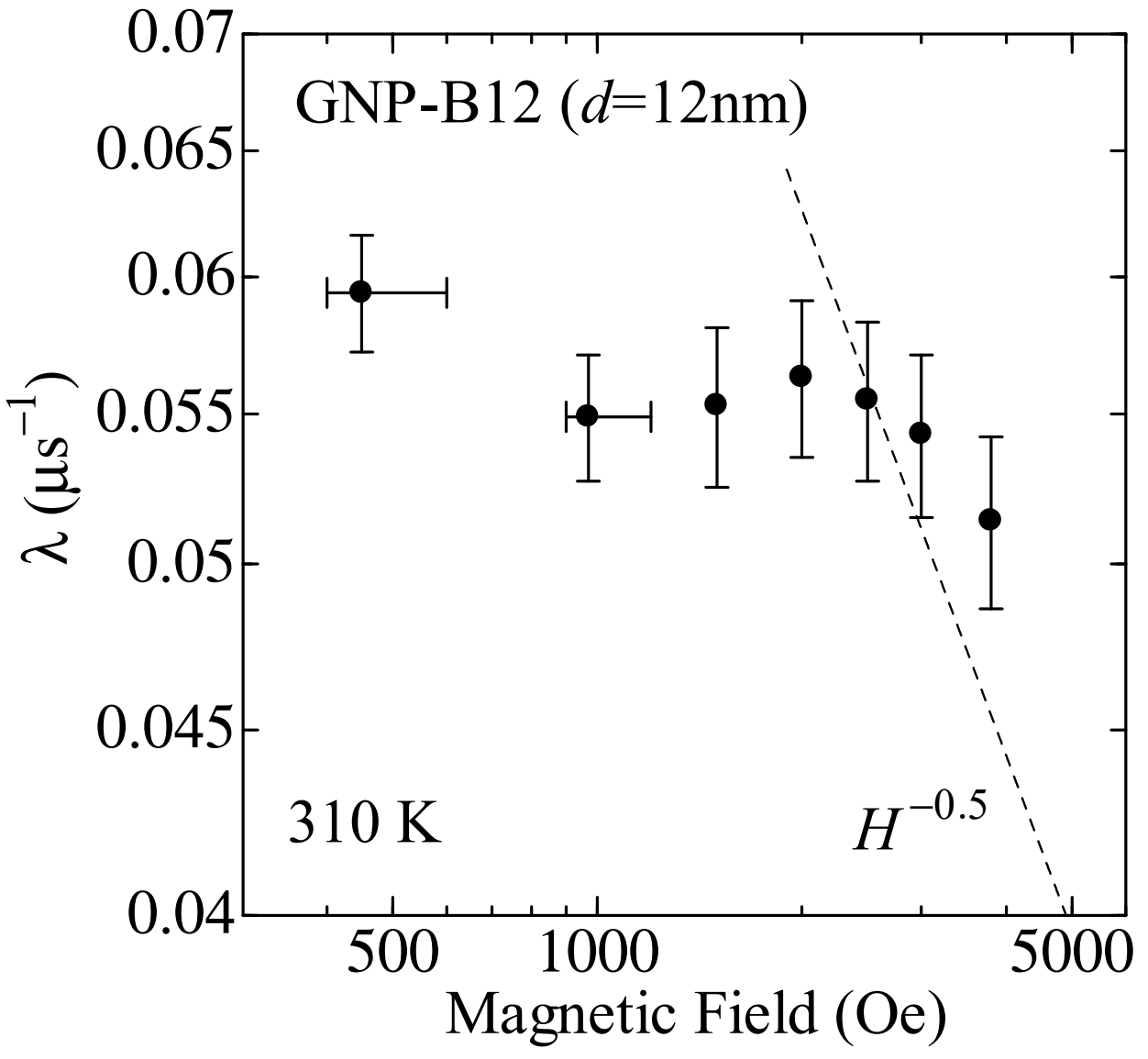

Fig. 3. LF-Field dependence of dynamical spin depolarization rate λ for B12 sample. Some data points on nearby fields are averaged to obtain high statistics. The dashed line shows the theoretical dependence $1/\sqrt{H_{\mathrm{LF}}}$ of 1D-spin-diffusion case.

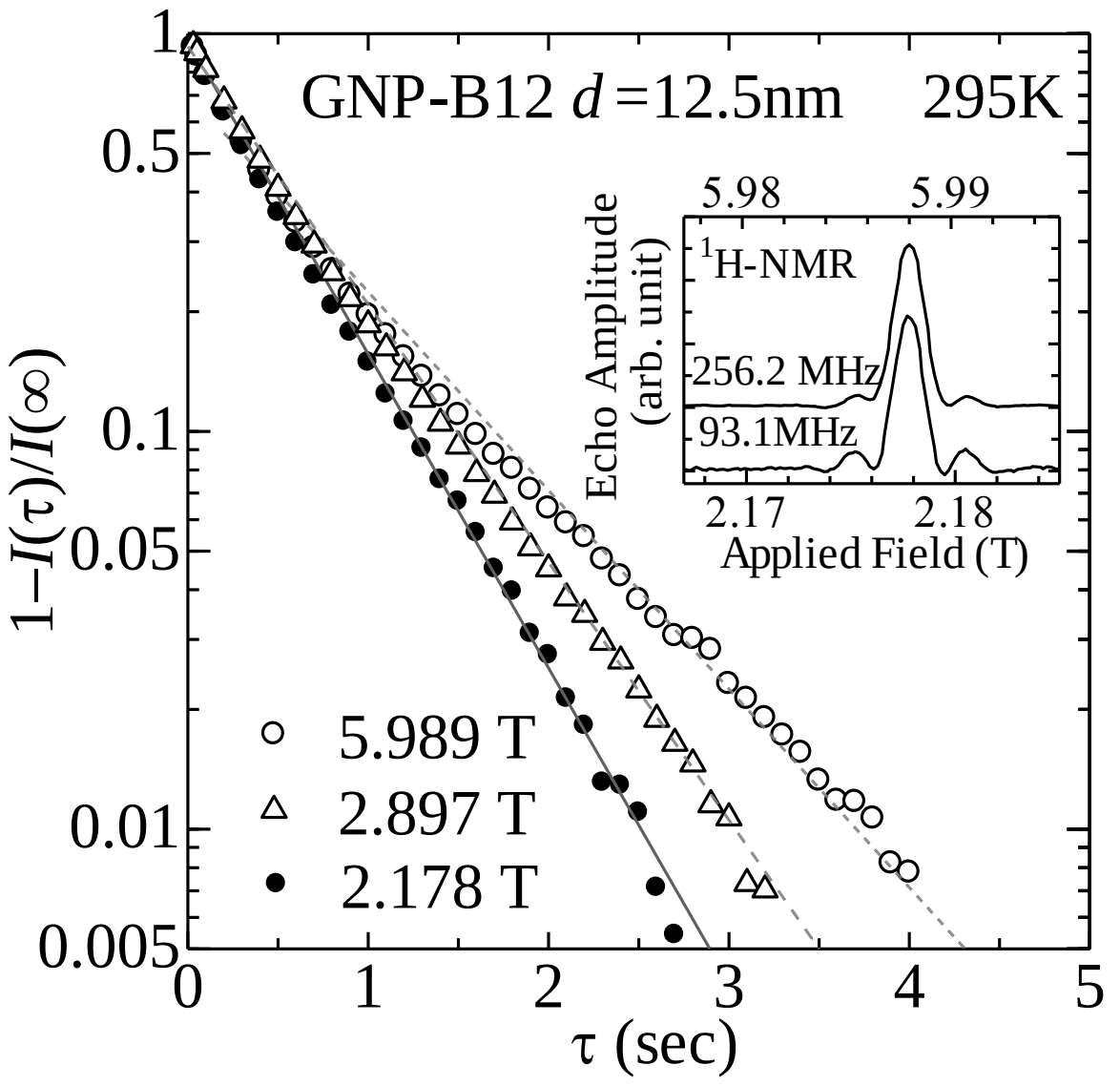

Fig. 4. Typical NMR-$T_1$ recovery curves of the B12 sample, measured at different fields. Dashed lines are exponential function fitted to data. The inset shows spectra taken under two different fields.

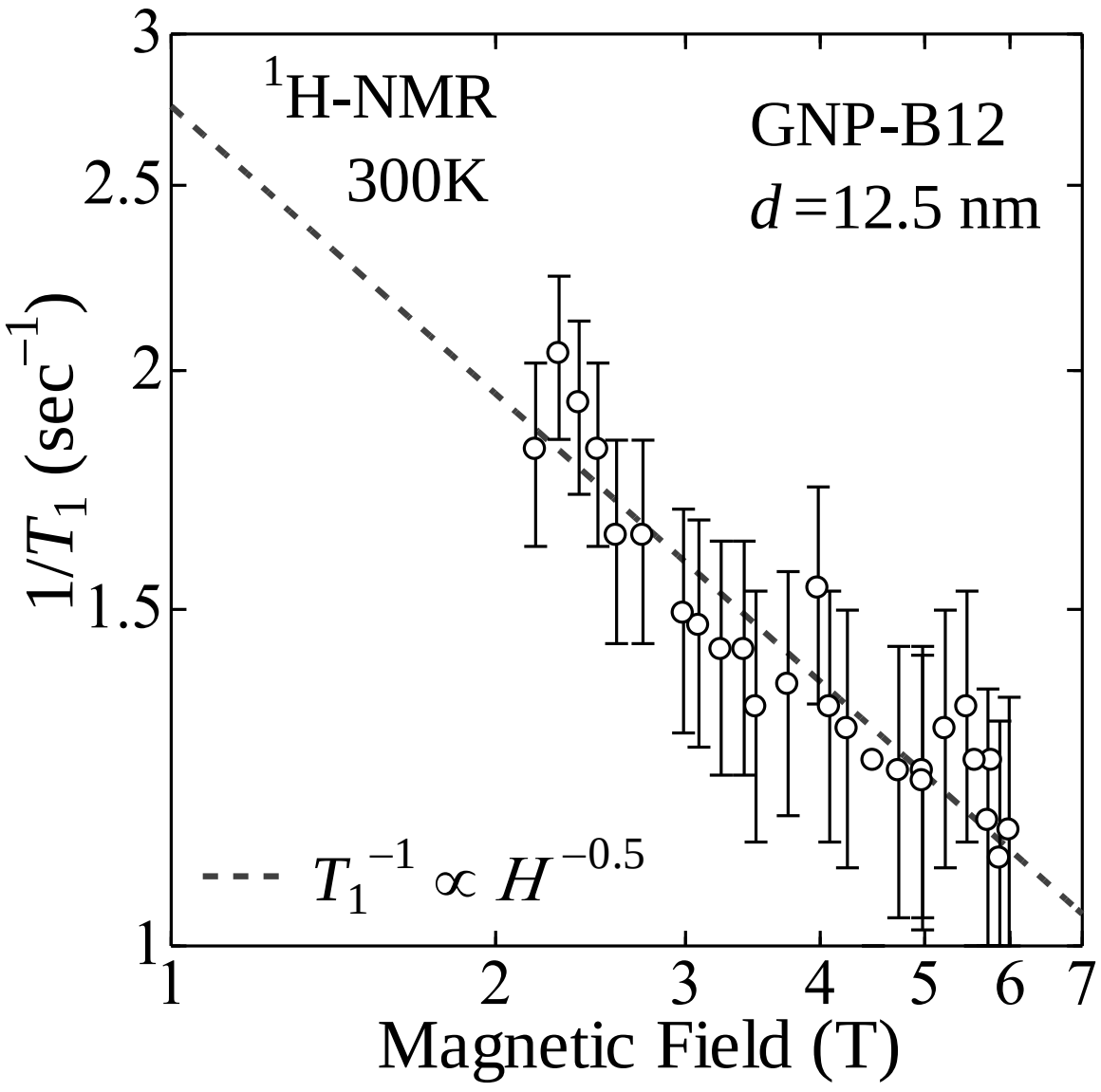

Fig. 5. Field dependence of NMR-$1/T_1$ for the sample B12. Dashed line shows the theoretical function for the 1D-spin diffusion case.

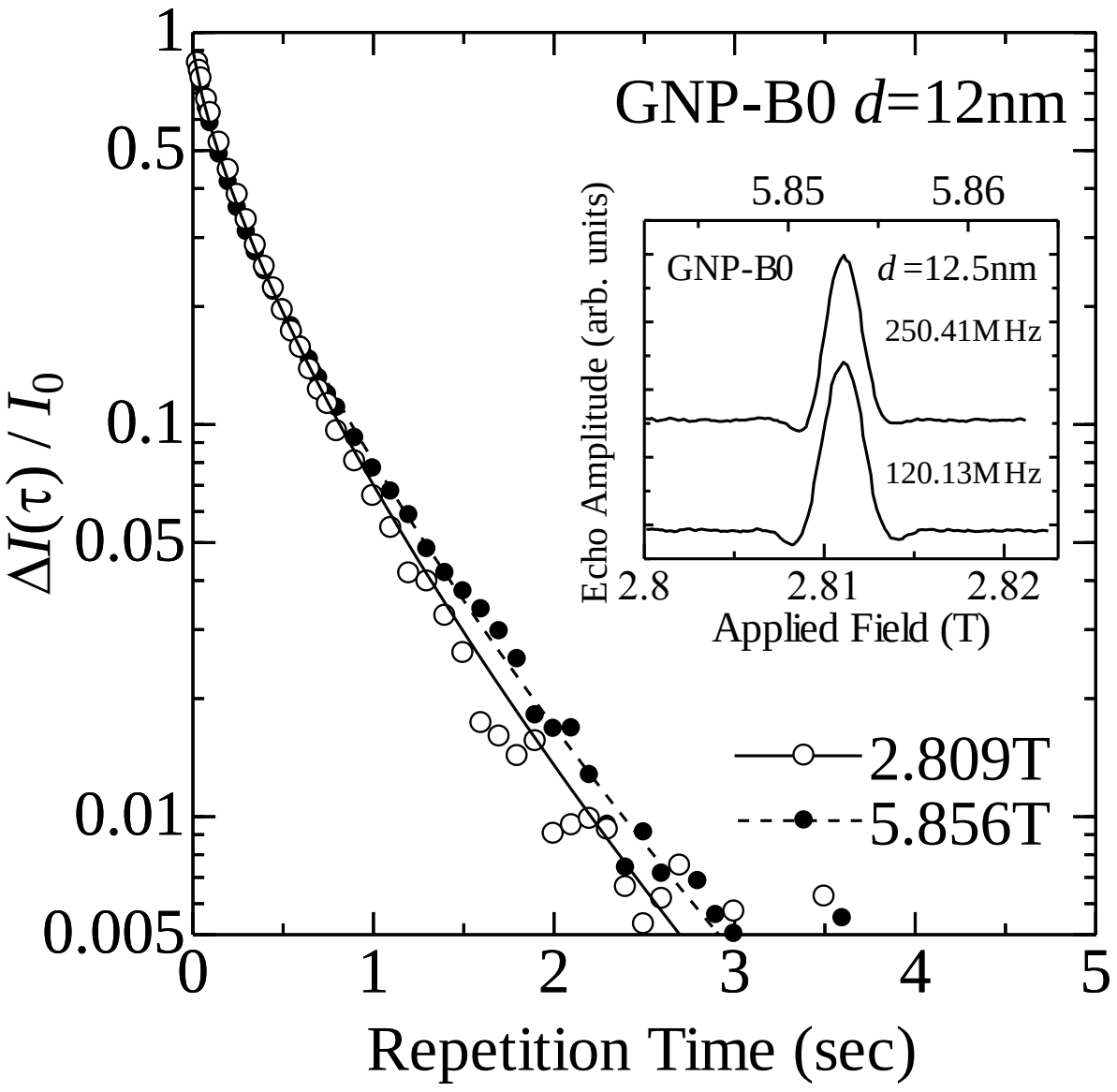

Fig. 6. Typical [1]H-NMR-$T_1$ recovery curves for the B0 sample taken at about 2.8 and 5.8 T. The dashed and solid curves are the stretched exponential function with $\beta =$ 0.7 fitted to data. The inset show spectra measured under the two fields.